\documentclass[aps,prb,twocolumn,superscriptaddress,showpacs]{revtex4}
\usepackage{graphicx}

\begin{document}

\preprint{APR 2003-XX}

\title{Resonant microwave properties of a voltage-biased
single-Cooper-pair transistor}

\author{L. Y. Gorelik}
\email{gorelik@fy.chalmers.se}
\affiliation{Department of Applied
Physics, Chalmers University of Technology and G\"{o}teborg
University, SE-412 96 G\"{o}teborg, Sweden}
\author{S. I. Kulinich}
\affiliation{Department of Applied Physics, Chalmers University of
Technology and G\"{o}teborg University, SE-412 96 G\"{o}teborg,
Sweden} \affiliation{B.~I.~Verkin Institute for Low Temperature
Physics and Engineering, 47 Lenin Avenue, 61103 Kharkov, Ukraine}
\author{R. I. Shekhter}
\affiliation{Department of Applied Physics, Chalmers University of
Technology and G\"{o}teborg University, SE-412 96 G\"{o}teborg,
Sweden}
\author{M. Jonson}
\affiliation{Department of Applied Physics, Chalmers University of
Technology and G\"{o}teborg University, SE-412 96 G\"{o}teborg,
Sweden}

\date{\today}

\begin{abstract}
We consider the microwave dynamics and transport properties of a
voltage-biased single-Cooper-pair transistor. The dynamics is
shown to be strongly affected by interference between multiple
microwave-induced inter-level transitions. As a result the
magnitude and direction of the dc Josephson current are extremely
sensitive to small variations of the bias voltage and to changes
in the frequency of the microwave field.
\end{abstract}


\maketitle

\section{Introduction\label{Introduction}}

It has long been known that transport properties of a
superconducting weak link may be strongly affected by external
time dependent (ac) fields. To give an example, just after the
discovery of the Josephson effect it was found that an alternating
rf-field will rectify the ac Josephson current that results from a
dc voltage bias and produce voltage steps --- Shapiro steps --- in
the current-voltage characteristics \cite{shapiro}. This effect
can be viewed as a synchronization of the Josephson current
oscillations of frequency $\omega_J=2eV/\hbar$, generated by a
bias voltage $V$, and the current oscillations induced by the ac
field. If the Josephson coupling energy $E_J$ is a simple harmonic
function of the superconducting phase difference $\Phi$, $E_J
(\Phi)=E_J\cos\Phi$, the positions $V^{(n)}$ of the Shapiro steps
are given by the expression $eV^{(n)}=n\hbar\omega$
($n=1,2,\ldots$), where $\omega$ is the frequency of the ac field.
Deviations from a simple harmonic relation between $E_J$ and
$\Phi$ result in additional "fractional" steps at voltages
$eV^{(nm)}=n\hbar\omega/m$. However their heights are
exponentially small when $m \gg n$.

\begin{figure}
\includegraphics[width=8cm]{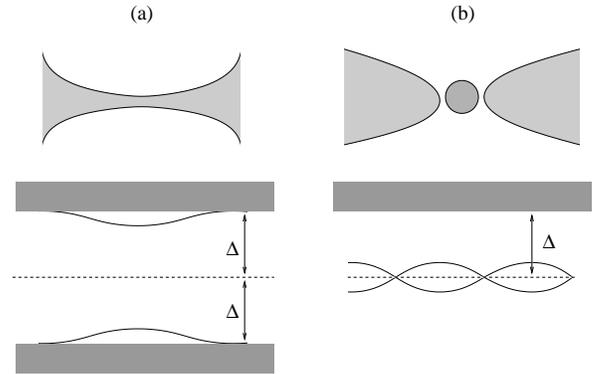}
\caption{ Schematic energy spectrum of (a) a superconducting point
contact (SPC) and (b) a single-Cooper-pair transistor (SCPT).
Energies are plotted as a function of the superconducting phase
difference $\Phi$ across the system. An important difference
between these two examples of a mesoscopic weak link is that the
discrete energy levels inside the gap $\Delta$ are well separated
from the continuum levels (dark bands) for an SCPT but not for an
SPC (see text). } \label{Fig1}
\end{figure}

New opportunities to influence the transport properties of
Josephson junctions by external ac fields appear in mesoscopic
systems, where only few quantum states are responsible for the
superconducting charge transfer. For such weak links the Josephson
energy --- no longer a macroscopic quantity (being proportional to
the area of the junction) --- may be of the same order of
magnitude as the energy quantum $\hbar \omega$ of the external ac
field. If so, the alternating field may significantly affect the
current by inducing resonant inter-level transitions between
charge carrying states in the junction and thereby changing their
population. The internal structure of the mesoscopic weak links
determines the physical origin of the discrete energy spectrum and
the relative positions of the energy levels inside the
superconducting gap $2\Delta$.

Superconducting point contacts (SPCs) and single-Cooper-pair
transistors (SCPTs) are two fundamentally different examples of
mesoscopic weak links. In an SPC such as the one sketched in
Fig.~\ref{Fig1}a, the rapid variation of the superconducting phase
across the junction gives rise to quasiparticle bound states
(Andreev states) that are localized to the vicinity of the
contact. The discrete energy levels (Andreev levels) associated
with these states appear inside the superconducting gap as shown
in Fig.~\ref{Fig1}a. Their positions $\epsilon_{i}$ are defined by
the single electron scattering matrix and in the case of low
transparency contacts one finds that $\Delta
-|\epsilon_i|\ll\Delta$. In an SCPT, which is composed of two
low-transparency tunnel junctions in series forming a small island
as sketched in Fig.~\ref{Fig1}b, the Coulomb blockade phenomenon
\cite{shekh} strongly affects the structure of the energy
spectrum. If the superconducting gap $\Delta$ is greater than the
Coulomb energy of a grain charged by a single electron, the ground
state of the SCPT with a completely isolated central dot, for
certain gate voltages, is degenerate with respect to adding one
more Cooper pair to the dot \cite{gms, mat}. Weak Cooper-pair
tunnelling that removes this degeneracy gives rise to current
carrying Cooper-pair resonant states, separated --- as illustrated
in Fig.~\ref{Fig1}b --- from the continuum spectrum by an energy
of order $\Delta$ \cite{nct, sym, dev, gor, ave}.

There is a qualitative difference between an SPC and an SCPT in
how the energy spectrum depends on the superconducting phase
change across the system. In an SPC the subgap level (as described
above this is a discrete energy level in the gap between the
ground state and the continuum excitation spectrum) is mixed with
the continuum spectrum at certain values of the phase difference,
while the Cooper-pair resonant levels in an SCPT are always
separated from the continuum, forming a well defined two-level
system. The microwave dynamics of a biased SPC was explored in
Ref.~\onlinecite{lundin}, where it was shown that quantum
interference among a sequence of temporally localized inter-level
transitions results in nontrivial transport properties of a
Josephson weak link of this type. In an SPC, however, coherence
may be preserved only during a fixed length of time, since the
discrete Andreev states die away as they periodically merge with
the continuum spectrum at certain values of the phase difference.
(We recall that a voltage bias makes the phase difference increase
linearly with time). In contrast, since the discrete energy levels
are here always separated from the continuum, the
microwave-induced transport properties of a biased SCPT is to a
great extent defined by the coherent dynamics of the two-level
system and strongly affected by interference processes. In order
to demonstrate this phenomenon, we explore below the microwave
dynamics and transport properties of a biased SCPT.

\begin{figure}
\includegraphics[width=8cm]{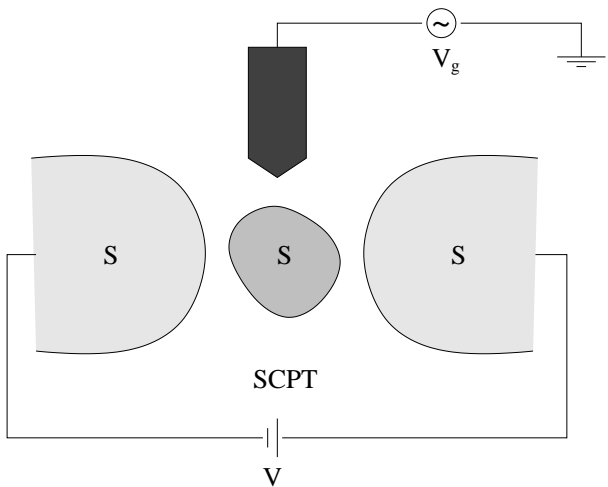}
\caption{ Sketch of a single-Cooper-pair transistor (SCPT)
composed of two low-transparency tunnel junctions in series
forming a small superconducting (S) island bridging the gap
between two superconducting bulk leads. The potential of the
island is controlled by a voltage $V_g$ applied to the gate
electrode, which is capacitively coupled to the island. A bias
voltage $V$ is applied across the SCPT giving rise to a linear
time dependence of the phase difference $\Phi = \omega_{J}t\equiv
2eV t/\hbar$ between the leads} \label{Fig2}
\end{figure}

\section{Model\label{Model}}
A sketch of the SCPT structure to be considered here is presented
in Fig.~\ref{Fig2}, where a nanoscale superconducting island ---
capacitively coupled to a gate electrode --- is shown to serve as
a bridge between two bulk superconducting leads. The Hamiltonian
for the system is
\begin{equation}\label{Hamiltonian}
\hat H = \frac{(2e\hat n + Q)^2} {2C} - E_{J}^{L}\cos(\Phi/2 -
\hat{\phi}) - E_{J}^{R}\cos(\Phi/2 + \hat{\phi})\,,
\end{equation}
where the Cooper-pair number operator\cite{n} $\hat{n}$  and the
phase operator $\hat{\phi}$ refer to the superconducting
condensate on the grain, $[\hat{\phi},\hat{n}] =i$; $E_{J}^{L(R)}$
is the Josephson coupling energy between the grain and the left
(right) lead, $C$ is the total mutual capacitance and $Q = C_g
V_g$ is an induced charge controlled by the gate voltage $V_g$.
The first term in the Hamiltonian governs the electrostatic energy
associated with the number of extra Cooper pairs on the grain,
while the second and third terms describe the Josephson coupling
between grain and leads. Below we consider the symmetric case for
which $E_{J}^{L} = E_{J}^{R} = E_{J}$. Therefore, the Josephson
part of the Hamiltonian simplifies to
$-2E_{J}\cos(\Phi/2)\cos\hat{\phi}$.

It is well known that if the characteristic Coulomb energy $E_{q}
\sim (2e)^2/2C$ associated with charge fluctuations due to Cooper
pair exchange between grain and leads, is significantly larger
than the Josephson energy $E_{J}$, while at the same time the
Coulomb energy $\sim e^2/2C$ related to {\it single} electron
fluctuations is smaller than the superconducting energy gap, the
low-energy states of the grain are given by a superposition of
only two charge states that differ in charge by 2e. Below we will
assume that the gate voltage is chosen in such a way that $Q=-e +
\delta Q$ where $|\delta Q| \ll e$. In that case the state of the
grain is a superposition of the charge state $|0\rangle$,
describing the neutral grain, and the charge state $|1\rangle$
with one extra Cooper pair. If $\delta Q = 0$ the Coulomb energy
in the states $|0\rangle$ and $|1\rangle$ are the same. This
degeneracy is, however, removed by the Josephson coupling that
induces Cooper pair exchange between the mesoscopic grain and the
bulk electrodes. As a result the eigenstates of the Hamiltonian
(\ref{Hamiltonian}) with eigenvalues $E_{\pm} = \mp
E_{J}\cos\Phi/2$ are given by symmetric and antisymmetric
superpositions of the states $|0\rangle$ and $|1\rangle: \vert \pm
\rangle=(\vert 0 \rangle \pm \vert 1 \rangle)/\sqrt 2$. These
states, which we will refer to as Cooper-pair resonant states,
will --- if occupied --- carry currents given by $j_{\pm}=\pm
eE_{J}\hbar^{-1}\sin\Phi/2$.

\begin{figure}
\includegraphics[width=8cm]{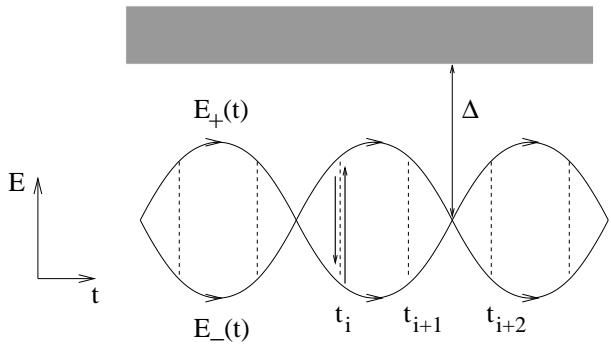}
\caption{ Time evolution of the Josephson levels $E_\pm (t)$ in
the single-Cooper-pair transistor (SCPT) of Fig.~\ref{Fig2}.
Dashed lines represent microwave-induced inter-level transitions
occurring at times $t_i$, when the resonance condition
$|E_+(t_i)-E_-(t_i)|=\hbar\omega$ is satisfied. } \label{Fig3}
\end{figure}

A bias voltage makes the phase difference over the junction change
with time, $\Phi(t) = \omega_J t+\Phi_0$. This time dependence
generates a periodic variation of the relative positions of the
Josephson levels as well as a periodic change in the current that
can be carried by the corresponding resonant Cooper pair states
(see Fig.~\ref{Fig3}). The total Josephson current through the
junction depends on the relative "population" $\delta \rho =
\rho_{+} - \rho_{-}$ of the levels and is given by the formula
$j_{J}=eE_{J}\hbar^{-1}\delta\rho\sin\Phi/2$ ($\rho_{\pm}$ are
probabilities to find the system in the states $|\pm\rangle$,
$\rho_++\rho_-=1$). One can find important differences between the
ac Josephson effect in such a system and in ordinary Josephson
junctions. If the Coulomb blockade is totally lifted, there are no
inter-level transitions as the levels cross each other and
consequently $\delta\rho$ is controlled only by relaxation
processes. Even if the temperature is much lower than the
characteristic inter-level distance, a weak relaxation to the
ground state with rate $ \nu \ll \omega_{J}$ leads to an
equalization of the level populations. As a result the Josephson
current will be proportional to $\nu$ and in the limit $\nu
\rightarrow 0$ not only the dc but also the ac current through the
Josephson junction vanishes.

The situation is different if an oscillatory voltage is applied to
the gate. A time dependent gate voltage makes the induced charge
vary in time as well, $\delta Q = \delta Q_{0}\cos(\omega t +
\varphi)$. This in turn gives rise to inter-level transitions
which control the level populations and by this means one may
stimulate the Josephson current. The most interesting case, which
will be considered below, appears when the frequency of the gate
voltage oscillations $\omega$ is of the same order as the
frequency of charge fluctuations on the grain $E_{J}/\hbar$. In
this case the inter-level transitions have a resonant nature and
significantly influence the Cooper pair transport even for
relatively small variations of the gate potential.

In the present work we consider a situation where the energy
quantum $\hbar\omega$ of gate voltage oscillations, is of the same
order but smaller than the maximum inter-level energy distance
$2E_{J}/\hbar$. At the same time we take $\omega$ to be much
larger than the rate of the gate-voltage induced resonant
inter-level transitions, $e\delta Q_{0}/\hbar C$. Under these
conditions and if one considers the time evolution of the system
at low bias voltages, $\omega_J\ll \omega$, one can neglect
inter-level transitions except during short time intervals $\delta
t_{i}$ near the times $t_i$, when the resonance
condition\cite{not} $\hbar\omega = \pm 2E_{J}\sin \Omega t_{i}$ is
fulfilled ($\Omega= \omega_J/2$). During these time intervals
Landau-Zener type inter-level transitions will occur (see
Fig.~\ref{Fig3}). Therefore the evolution of the system we are
interested in is given by a sequence of scattering events
separated by periods of "free" evolution, during which the level
populations do not change.

Despite the fact that the level positions vary periodically in
time with period $T_0 = 2\pi/\Omega$, the dynamics of the system,
strictly speaking, is not periodic. This is because the phases of
the complex elements of the scattering matrix depend on the phase
$\omega t_{i}+ \varphi$ which defines the value of induced charge
at time $t_{i}$. This fact is of great importance for the
transport properties of the system. As we will show below, if the
ratio $\omega/\Omega$ is an irrational number the dynamics of the
system is qusiperodic and the dc Josephson current is equal to
zero. But if this ratio is a rational number, then a dc Josephson
current does flow through the junction. The maximum value of the
current appears when $\omega/\Omega$ is an integer.

\section{Theory}
Using the Cooper-pair resonant states representation one can
recast the Hamiltonian (\ref{Hamiltonian}) in the form
\begin{equation} \label{ham}
\hat H(t)=E_J\sigma_3 \sin\Omega t +\varepsilon\sigma_1 \cos
(\omega t + \varphi)\,,
\end{equation}
where $\varepsilon= 2e\delta Q_0/C$  and $\sigma_i$ are the Pauli
matrices. As already anticipated, we will assume that the
following relations are fulfilled:
\begin{eqnarray} \label{09}
&&\varepsilon, \hbar \Omega \ll \hbar \omega \leq E_J\ll \Delta\,,\\
&&\omega /\Omega = N + p/q\,, p < q\,, N\gg 1.\nonumber
\end{eqnarray}
Here $N,p,$ and $q$ are integer numbers. The final result, other
than details of calculations, does not depend in any essential way
on the arithmetic properties of $N, p, q$, and for definiteness we
take $q$ to be an even number ($p$ is odd). Under these conditions
the Hamiltonian (\ref{ham}) is a periodic function of time with
period $T=qT_0$.

The dc Josephson current is given by expression
\begin{equation} \label{12}
I=\frac{1}{T}\int_0 ^T dt \,\text{Tr}\, \hat \rho (t) \hat \jmath
(t)= \frac{2}{T}\int_0^{T/2} dt \, \text{Tr}\,\hat\rho (t)
\hat\jmath (t)\,,
\end{equation}
where $$\hat{\jmath} (t)=\frac{2e}{\hbar}\frac{\partial \hat H
(t)}{\partial \Phi}=\frac{e E_J}{\hbar}\sigma_3 \cos \Omega t$$ is
the current operator, $\hat\rho (t)$ is the density matrix of the
two-level system, while its time evolution is governed by the
Liouville-von Neumann equation
\begin{equation} \label{13}
\imath\hbar\frac{\partial \hat\rho}{\partial t}= [ \hat H, \hat
\rho]- \imath \hbar \nu (\hat \rho - \hat \rho_{eq})\,.
\end{equation}
Here $\hat \rho_{eq}(t)=(1-\sigma_3 \tanh(\beta E_J\sin\Omega
t))/2$ is the quasistatic density matrix for the unperturbed
Hamiltonian ($\varepsilon = 0$), $\beta$ is the inverse
temperature, $\nu$ is the relaxation rate. The formal solution of
Eq.~(\ref{13}) can be expressed as
\begin{equation} \label{16}
\hat\rho (t) = \nu \int_{-\infty}^t dt'\, e^{\nu (t'-t)} \hat
u(t,t')\hat \rho_{eq} (t') \hat u^{\dag}(t,t')\,,
\end{equation}
where
$$\hat u(t_2,t_1)=\hat T \text{exp}\left[
-\frac{\imath}{\hbar}\int_{t_1}^{t_2} dt \hat H(t) \right] $$ is
an evolution operator possessing the symmetry properties
\begin{equation} \label{11} \hat u (t_2 + q T_0/2, t_1
+qT_0/2)=\sigma_3
\hat u (t_2,t_1)\sigma_3 \,,
\end{equation}
which derives from the symmetry of the Hamiltonian (\ref{ham}),
$$ \hat H(t+T/2)=\sigma_3\hat H(t) \sigma_3 \,.$$
It follows that the density matrix $\rho (t)$ satisfies the
relation
\begin{equation} \label{17}
\hat \rho (t+T/2)=\sigma_3 \hat \rho (t) \sigma_3 \,,
\end{equation}
which we used in Eq.~(\ref{12}). In the limit $\nu T \ll 1$ this
relation together with Eq.~(\ref{16}) immediately results in the
equation
\begin{equation} \label{18}
\hat \rho _0 = (1-\tilde\nu) \,\hat U \hat \rho_0 \hat U^\dag
+\tilde\nu \hat F\,,
\end{equation}
for the density matrix $\hat \rho _0 =\hat \rho (t=0)$. Here
$\tilde\nu =\nu T/2 \ll 1$ and
\begin{eqnarray}
&\hat U&=\sigma_3 \hat u(qT_0/2,0)\,,\label{19}
 \\ \label{19a}
\hat F = \frac{2}{T}\int_0^{T/2}&dt& \sigma_3\hat u(T/2,t)
\hat\rho_{eqv}(t) \hat  u^\dag (T/2,t)\sigma_3\,.
\end{eqnarray}
Equation (\ref{18}) and the spectral properties of the unitary
operator $\hat U$ will play a central role as we proceed.

In the limit $\tilde\nu =\nu T/2 \ll 1$ the solution of
Eq.(\ref{18}) has the form
$$\hat \rho_0 =\hat I/2 + B (\vert \lambda_1\rangle\langle\lambda_1
\vert - \vert\lambda_2\rangle\langle\lambda_2\vert)\ +
{\mathcal{O}}(\tilde\nu)\,,$$ where $\vert\lambda_{k=1,2}\rangle$
are the eigenvectors of the operator $\hat U$ with eigenvalues
$\text{exp}(i\lambda_k)$ and $B = (\langle\lambda_1\vert \hat F
\vert \lambda_1\rangle -\langle\lambda_2 \vert\hat F \vert
\lambda_2 \rangle)/2\,.$ From the symmetry of the Hamiltonian
(\ref{ham}) it follows that $\sigma_2 \hat H(t) \sigma_2 = -\hat H
(t)$, $\vert \lambda_2\rangle=\sigma_2\vert\bar{\lambda}_1\rangle$
and $\lambda_1+ \lambda_2=\pi$. The expression $\hbar \lambda_k/T$
gives the quasienergies of the periodic Hamiltonian, $\hat H(t)$;
we will refer to $\lambda_k$ as a quasienergy factor.

The average Josephson current, Eq.(\ref{12}), can now be cast in
the form
\begin{eqnarray} \label{20}
&&I= B \sum_k  (-1)^{k+1} \langle\lambda_k\vert \hat I_0
\vert \lambda_k\rangle \,,\\
&&\hat I_0= \frac{2eE_J}{\hbar T}\int_0^{T/2}dt\cos \Omega t \,
\hat u^\dag (t,0) \sigma_3 \hat u(t,0)\,.
\end{eqnarray}
In order to calculate the population coefficient $B$ and the
average current in the quasienergy states, $\langle\lambda_k\vert
\hat I_0 \vert \lambda_k\rangle$, it is convenient to introduce an
effective Hamiltonian,  $\hat H(t;\xi)$, of the form
\begin{equation} \label{255}
\hat H(t;\xi)=\hat H(t)+ \xi \sigma_3\tanh (\beta E_J\sin \Omega
t) \,.
\end{equation}
One can show that Eq.~(\ref{20}) for the Josephson current can be
expressed as
\begin{equation} \label{112}
I=\frac{2e\hbar\omega}{\pi q T}\lim_{\xi\to 0} \frac{\partial
\lambda_1} {\partial \varphi}\,\frac{\partial \lambda_1}{\partial
\xi}\,,
\end{equation}
where $\lambda_1(\varphi;\xi)$ is the appropriate quasienergy
factor of the effective Hamiltonian (\ref{255}).

Only resonant inter-level transitions are of importance if the
microwave interaction energy, $\varepsilon$, is much smaller than
the characteristic inter-level distance $2E_J\sim\hbar\omega$.
Ignoring the nonresonant part of the Hamiltonian $\hat H(t;\xi)$
({\it i.e.} using the resonance approximation) one finds that the
evolution operator $\hat u((n+1)T_0/2, nT_0/2)$ has the form
\begin{eqnarray} \label{122}
&&\hat u((n+1)T_0/2, nT_0/2;\xi)= \sigma_1^n e^{ -i((n+1)\omega
T_0/2+ \varphi)\sigma_3/2}\nonumber\\
&&\quad\quad\times\,\hat u_{res}(T_0/2,0;\xi) \, e^{ i(n\omega
T_0/2+\varphi)\sigma_3/2}\sigma_1^n\,.
\end{eqnarray}
Here $\hat u_{res}(T_0/2,0;\xi)$ is the evolution operator
generated by the Hamiltonian $\hat H_{res}(t;\xi)$, where
\begin{eqnarray} \label{432}
\hat H_{res} (t;\xi)=(E_J \sin\Omega t -\hbar \omega/2&+&
\nonumber
\\\xi \tanh (\beta E_J \sin\Omega  t) )\sigma_3 &+& \varepsilon
\sigma_1 \,.\nonumber
\end{eqnarray}
From the symmetry of the Hamiltonian $\hat H_{res}(t;\xi)$ it
follows that the operator $\hat u_{res}(T_0/2,0;\xi)$ has the form
\begin{eqnarray} \label{433}
\hat u_{res}(T_0/2,0;\xi)&=& \nonumber \\
e^{i\theta\sigma_3}(\sqrt{1-\tau^2}+i\tau\sigma_1)
e^{i\theta\sigma_3} &\equiv& e^{i\theta\sigma_3}\sigma_1 \hat L
e^{i\theta\sigma_3}\,.\nonumber
\end{eqnarray}
The parameter  $\tau$ is the resulting probability amplitude for
an inter-level transition when the superconducting phase
difference changes from $-\pi$ to $0$.

Equation~(\ref{122}) permits us to represent the desired evolution
operator $\hat U$ of Eq.~(\ref{19}) as
\begin{eqnarray}  \label{124}
&&\hat U(\xi)= \sigma_3 \prod_{n=0}^{q-1}\hat u((n+1)T_0/2,
nT_0/2;\xi)=\\&& \pm i e^{-i\sigma_3(\theta-\varphi/2)}\left[
\prod_{n=0}^{q-1}\left( \hat L e^ {i(\pi\alpha n
+\varphi)\sigma_3} \right) \right] e^{i\sigma_3(\theta-\varphi/2)}
\,, \nonumber
\end{eqnarray}
where $\alpha\equiv p/q$ and the sign ($\pm$) depend on the
arithmetic properties of the natural numbers $N,p$,and $q$.

Now one can see that the eigenvalues of the unitary operator $\hat
U(\xi)$ can be found  by solving the difference equation
\begin{equation} \label{30}
\vert \psi(n+1)\rangle = \hat L e^{i (\pi \alpha n +
\varphi)\sigma_3}\vert \psi(n)\rangle \,
\end{equation}
with the boundary condition $\vert \psi (q)\rangle =
e^{i\lambda_k}\vert \psi (0)\rangle$, $k=1, 2.$ A "quasiperiodic"
equation of this type may be analyzed by method developed in
Ref.~\onlinecite{ggksj}. The result for the quasienergies
$\lambda_k$ from such an analysis is
\begin{equation} \label{02}
\lambda_k = (-1)^k\arccos (\tau ^q \cos q\varphi) + c_k(p,q,N) \,,
\end{equation}
where the constants $c_k$ are irrelevant.

Taking into account Eqs.~(\ref{112}) and (\ref{02}) the average
current can now be rewritten as
\begin{equation}
I= -\frac{e\hbar \omega}{2\pi T_0} \frac{\tau^{2(q-1)}\sin
2q\varphi}{1-\tau^{2q}\cos^2 q\varphi} \left( \frac{\partial
\tau^2}{\partial \xi} \right)_{\xi=0}\,.
\end{equation}

In this work we restrict ourselves to the low temperature limit,
where $\beta E_J\gg 1$. In this limit $ \hat H_{res}(t;\xi)= \hat
H_{res}(t;\xi-\hbar \omega/2)$ and, as a consequence, $ \partial
\tau^2/\partial \xi =-2\hbar^{-1}\partial \tau^2/\partial \omega$.
Therefore, in order to calculate the current at low temperatures
one can use the Hamiltonian $\hat H_{res}(t;\xi=0)$.

As already mentioned, only resonant inter-level transitions are
important here. In the time interval $(0, T_0/2)$ the resonant
condition $2E_J\sin \Omega t_i=\hbar \omega$ is fulfilled only at
the two times $t_0= \Omega ^{-1} \arcsin \hbar \omega/2E_J$ and
$t_1=T_0/2-t_0.$ As these "points" in time are reached,
inter-level Landau-Zener transitions \cite{lz} (see
Fig.~\ref{Fig3}) occur with probability amplitude
$w^{1/2}e^{i\vartheta}$, where
$$ w=1-\text{exp}\, \left[ -\frac{\pi \varepsilon^2}{4E_J \hbar
\Omega}\left[1-\left( \frac{\hbar \omega}{2E_J}
\right)^2\right]^{- \frac{1}{2}}\right]\,.$$

Dividing time interval $(0,T_0/2)$ into periods of free evolution
(where interactions with the microwave field can be neglected) and
describing the passage through the resonant points by a
Landau-Zener scattering matrix, one finds the probability for an
inter-level transition to be
\begin{eqnarray}
&&\tau=2 w\sqrt{1-w^2}\cos(\Theta + \vartheta) \nonumber \\ &&
\Theta = \hbar^{-1}\int_{t_0}^{T_0/2-t_0}dt\, E_J\sin \Omega t
-\omega(T_0/4-t_0) \nonumber
\end{eqnarray}
and the final expression for the microwave-induced dc Josephson
current becomes
\begin{equation} \label{final}
I= \frac{e\omega}{\pi^2}\left( \arccos \frac{\hbar \omega}{2E_J}
\right) \frac{\tau^{2q}\tan(\Theta+\vartheta)}{1-\tau^{2q}\cos^2
q\varphi} \sin 2q\varphi  \,.
\end{equation}
Equation (\ref{final}) determines the current-voltage
characteristics of the SCPT studied, which depend crucially on the
frequency of the microwave irradiation. Its validity is restricted
to the special values of $\omega$ and $V$ determined by
Eq.~(\ref{09})

\begin{figure}
\includegraphics[width=8cm]{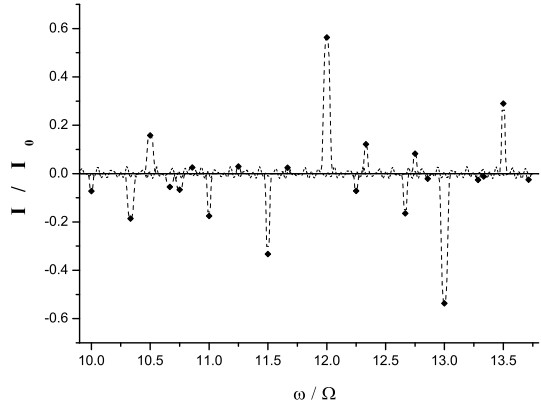}
\caption{ Microwave-induced current $I$ in units of $I_0\equiv
e\omega/\pi^2$, plotted as a function the microwave frequency
$\omega$ normalized to $\Omega\equiv\omega_J/2$. The result was
obtained with $\varphi=\pi/11,~ \hbar\omega/2 E_J=0.001,~ w=0,5$
and $q_{max}=7$ (see text).} \label{Fig4}
\end{figure}

Averaging the current over a long time $T$ (corresponding to a
large value of $q$, since $T=qT_0$), and plotting it as a function
of the parameter $\omega/\Omega$, results in many sharp features.
(Notice that two values of $\omega/\Omega$ which are arbitrarily
close to each other might correspond to widely different values of
$q$ in Eq.(\ref{09})). It is clear, however, that any upper limit
$T_{max}$ for the allowed averaging time will tend to smooth out
the sharp features. The value of $T_{max}$ might be determined by
either a relaxation time or a finite observation time. In either
case currents calculated for two values of $(\omega/\Omega)_{1,2}$
will be indistinguishable if $|q_1-q_2|>\Omega T_{max}$. This
argument provides a justification for smoothening the current as
given by Eq.~(\ref{final}) in order to get a regular dependence
$I=I(\omega/\Omega)$. An appropriate procedure is to evaluate
Eq.~(\ref{final}) at all rational points for which $q<q_{max}$.
Then a smooth dependence $I(\omega/\Omega)$ can be obtained by
making use of an interpolation procedure. The result of such an
interpolation procedure is presented in Fig.~\ref{Fig4}.

\section{Conclusion}
We have shown that the microwave dynamics of a single-Cooper-pair
transistor is strongly influenced by interference effects caused
by the quantum dynamics of the superconducting dot, which --- as
in Fig.~\ref{Fig2} --- forms the central island in the transistor
structure. The sharp peaks that were found in the current-voltage
($I-V$) characteristics for "fractional" values of the voltage, as
shown in Fig.~\ref{Fig4}, are a signature of a resonant
interaction with the microwave field. The peaks occur when $eV =
\hbar\omega/(N + p/q)$, where $N$, $p$, and $q$ are integers; they
are finite if $N \rightarrow \infty$ and do not vanish in the
limit of weak microwave irradiation, The interaction with the
microwave field results in multiple, coherent transitions between
resonant Cooper-pair levels. This picture, which differs
qualitatively from the Shapiro effect \cite{shapiro}, is a direct
manifestation of the role of strong Coulomb correlations in the
nonequilibrium superconducting dynamics of mesoscopic weak links.
The Coulomb blockade of Cooper pair tunnelling reduces the number
of relevant charge states to two, and is responsible for the
ensuing quantum coherent two-level dynamics of Cooper pair
resonant states.

A single-Cooper-pair transistor requires the Coulomb blockade
energy $E_{q}$ to be much larger than the Josephson coupling
energy $E_{J}$. If instead $E_{J} \gg E_{q}$, a large number of
dot charge states with different number of Cooper pairs are
degenerate. As a result tunnel coupling to the leads form dot
states with a well defined superconducting phase. Microwave
properties in this limit are governed by the {\em classical
dynamics} of the superconductiong phases, caused by the driving
voltage and the microvawe field. If we were to imagine that
$E_{q}$ could be diminished from a value much larger than $E_{J}$
to a value much smaller than $E_{J}$, we would witness a
transition from a situation where the quantum dynamics of the
superconducting phase was relevant to one where the classical
dynamics would dominate. During such a transition the resonance
microwave absorption would disappear and be replaced by standard
Shapiro signatures, known from macroscopic weak links.

\begin{acknowledgments}
This work has been supported financially by the Royal Swedish
Academy of Sciences, KVA, and by the Swedish Research Council, VR,
(LYG, RIS). SIK acknowledges the hospitality of the Department of
Applied Physics at Chalmers University of Technology and
G\"{o}teborg University.
\end{acknowledgments}


\begin{thebibliography}{0}
\bibitem{n} We count the excess Cooper pairs relative to the neutral
state.
   Hence the eigenstate of the operator $\hat{n}$ with eigenvalue
   $n=0$ corresponds to a neutral grain.
\bibitem{not} For the sake of definiteness we let $\Phi_0=-\pi$.
\end{thebibliography}
\end{document}